\documentclass[10pt,journal,compsoc]{IEEEtran}
\ifCLASSOPTIONcompsoc
  \usepackage[nocompress]{cite}
\else
  \usepackage{cite}
\fi
\usepackage{color}

\ifCLASSINFOpdf
  \usepackage[pdftex]{graphicx}
\else
\fi
\begin{document}
%
\title{Revisiting Walking-in-Place by Introducing Step-Height Control, Elastic Input, and Pseudo-Haptic Feedback}
%
%
%
%

\author{Yutaro~Hirao,~
        Takuji~Narumi,~
        Ferran~Argelaguet,~
        and~Anatole~Lécuyer 
\IEEEcompsocitemizethanks{\IEEEcompsocthanksitem Y. Hirao and T. Narumi are with the University of Tokyo, Tokyo 113-8654, Japan. E-mail: \{hirao, narumi\}@cyber.t.u-tokyo.ac.jp.
\IEEEcompsocthanksitem A. Lécuyer and F. Argelaguet work with Univ. Rennes, Inria, IRISA, CNRS, Rennes, France.}
\thanks{}}


%
%

\markboth{Journal of}
{Hirao \MakeLowercase{\textit{et al.}}: Revisiting Walking-in-Place}
%



\IEEEtitleabstractindextext{%
\begin{abstract}
Walking-in-place (WIP) is a locomotion technique that enables users to ``walk infinitely" through vast virtual environments using walking-like gestures within a limited physical space. This paper investigates alternative interaction schemes for WIP, addressing successively the control, input, and output of WIP. First, we introduce a novel height-based control to increase advanced speed. Second, we introduce a novel input system for WIP based on elastic and passive strips. Third, we introduce the use of pseudo-haptic feedback as a novel output for WIP meant to alter walking sensations. The results of a series of user studies show that height and frequency based control of WIP can facilitate higher virtual speed with greater efficacy and ease than in frequency-based WIP. Second, using an upward elastic input system can result in a stable virtual speed control, although excessively strong elastic forces may impact the usability and user experience. Finally, using a pseudo-haptic approach can improve the perceived realism of virtual slopes. Taken together, our results suggest that, for future VR applications, there is value in further research into the use of alternative interaction schemes for walking-in-place.


\end{abstract}

\begin{IEEEkeywords}
walking-in-place, pseudo-haptics, passive haptics, elastic input, virtual reality, locomotion.
\end{IEEEkeywords}}

\maketitle

\IEEEdisplaynontitleabstractindextext

%
\IEEEpeerreviewmaketitle

\IEEEraisesectionheading{\section{Introduction}\label{sec:introduction}}

%
%
%
%
\IEEEPARstart{T}{he} development of virtual reality (VR) technologies enables the creation of vast virtual environments (VEs) without physical limitations. However, the user's physical motion is restricted by the available tracking space. One of the promising approaches to tackle this problem is the walking-in-place (WIP) technique, which utilizes in-situ walking-like gestures \cite{nilsson2016walking, Nilsson:2018}. 
WIP techniques are achieved with relatively inexpensive systems \cite{Feasel:2008} and expected to improve the sense of presence in VR compared to in-situ locomotion, such as using hand-held controllers \cite{Slater:1995, bhandari2017legomotion}. However, the WIP technique has several problems such as a small range of stably attainable walking speeds and lacking cost-effective methods to simulate various VEs. Therefore, we revisited WIP to address these problems.

\begin{figure}[t]
 \centering
 \includegraphics[width=\columnwidth]{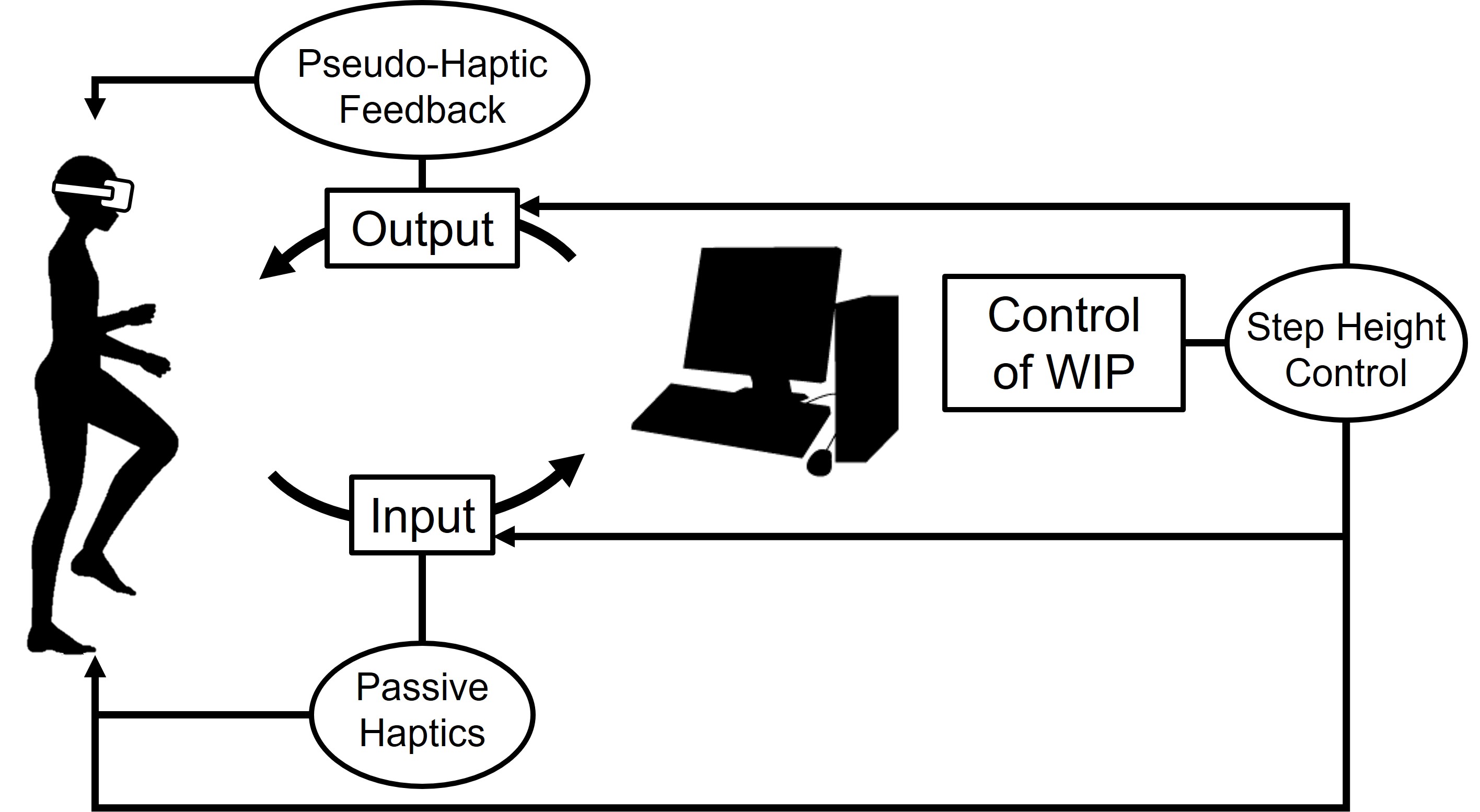}
 \caption{General scheme of a walking-in-place (WIP) system with our main contributions. Three factors have been investigated: control, input, and output of WIP - enclosed in squares. The novel approaches introduced in this study are enclosed in circles.}
 \label{overview}
\end{figure}

Fig.\ref{overview} shows the general interaction scheme of a WIP system and the additional factors we investigated in this paper. In this paper, we revisited three factors of this scheme, considering their relationships and synergies. 

First, the control factor is related to the ability of the user to control the virtual speed, that is, the algorithm of WIP. This paper introduces a novel control algorithm that relies on step height together with the conventional step frequency. We hypothesized that the step height as a second parameter to adjust the walking speed improves the range of stably attainable virtual speeds.

Second, the input factor is related to the input device or method used in the WIP system. This study investigates a novel passive and elastic input system. Elastic input is cost-effective and it is known that passive haptics such as elastic input can improve the task performance and realism of VEs \cite{Paljic:2004, paljic2004evaluation, Achibet:2014, Achibet:2015}. In addition, Considering the synergy of step height control and elastic input, we hypothesized that the elastic input provides users with a better sense of their step height and leads to better control of the WIP speed than without the elastic input. Here, the elastic force might even be considered as the output of the system. We think that the elastic force can be considered as both output and input, and in this paper, we considered it as an input factor, following the context of HCI studies \cite{zhai1993human, lecuyer2009simulating, paljic2004evaluation}.

Third, the output factor represents the type of feedback sent by the WIP system to the user based on the user's stepping gestures. This paper uses a pseudo-haptic approach \cite{Pusch:2011} to simulate body load sensations such as the one experienced when walking on a slope by adjusting the output speed gain to the stepping gesture to enhance the realism of the VE. The pseudo-haptic approach has a synergy with the elastic input and the combination of them improves the realism of the interaction with the VE \cite{Paljic:2004, paljic2004evaluation, Achibet:2014, Achibet:2015}. Furthermore, our step height and frequency control would reflect users' behavioral changes induced by the pseudo-haptic feedback to the output speed better than step frequency control only. 

This study investigates the impact of these three new interaction schemes on WIP performance and user experience (UX), and their synergy through three experiments. In Section 2, we introduce related research in terms of each WIP interaction scheme. In Section 3, we present an experiment that examines the effects of step height and frequency (SHeF) control on users' WIP control strategies and WIP performance. In Section 4, we describe an experiment on the effects of SHeF control and elastic input systems on the performance and UX of the WIP. In Section 5, we describe experiments investigating the effects of pseudo-haptic feedback for a virtual slope representation, with SHeF control and elastic input, on the realism of the VE and UX.

\section{Related Work}

In this section, we introduce related research in terms of each WIP interaction scheme.

\subsection{Input of WIP}
To improve the realism of walking in VEs, many techniques for recognizing users’ walking-like gestures have been studied. Walking-like gestures include arm swinging \cite{Nilsson:2013}, head shaking \cite{Terziman:2010}, feet sliding \cite{CyberShoes,Templeman:2007}, heal tapping \cite{nilsson2013tapping}, leaning of the torso \cite{wang2018step}, walking on a treadmill/interface \cite{Darken:1997,Omni,Iwata:2001,Iwata2005}, and stepping gestures. Other work has enabled users to create their own gestures \cite{kim2021user}. In this paper, we focused on the WIP technique using stepping gestures; which is related to real walking more closely than other gestures; only requires tracking of the user’s gestures without any additional equipment; and has a synergy with passive elastic force.


Swanson \cite{Swanson:2003} defined passive haptic interfaces as interfaces using ``energetically passive actuators, which may in general only remove, store, or redirect kinetic energy within the system." Elastic force is one of the major passive haptic forces and is considered as an important factor of the input devices in the field of HCI and pseudo-haptics studies. For example, Zhai and Milgram compared isometric and elastic rate controllers, and confirmed that the elastic controller improves users’ performance \cite{zhai1993human}. Moreover, several pseudo-haptics studies confirmed that elastic input devices are cost-effective and can improve the task performance and realism of VEs \cite{Paljic:2004, paljic2004evaluation, Achibet:2014, Achibet:2015}. Treadport is a device that uses passive elastic force with WIP technique \cite{Hejrati:2015}. It can present a horizontal elastic force to improve the realism of walking and WIP control on a treadmill. However, Treadport requires a relatively complex device and is intended to be used with a treadmill. In this paper, we investigate the way to introduce elastic input to WIP in a more cost-effective way.


\subsection{Control of WIP}
Several algorithms for controlling virtual walking speed by stepping gesture recognition have been studied. Templeman et al. \cite{Templeman:1999} proposed Gaiter, which is a WIP technique that enables the control of speed and direction using pressure sensors on the sole and knee trajectories. Feasel et al. \cite{Feasel:2008} proposed low-latency, continuous-motion (LLCM) WIP, which was based on the speed of the user’s heel motion while walking in place. Wendt et al. \cite{Wendt:2010} proposed gait-understanding-driven (GUD) WIP, which utilized the biomechanics of the human gait to understand the stepping gestures and calculated the virtual speed using step frequency. Because GUD-WIP is based on an understanding of the human gait biomechanics, it can create a more consistent walking speed at the same frequency as walking compared to LLCM-WIP. Furthermore, Bruno et al. \cite{Bruno:2013} proposed a speed-amplitude-supported WIP that created a virtual speed output using foot amplitude and velocity, which they found were positively correlated. A more recent WIP study has used convolutional neural networks for improved WIP control, such as starting, stopping, and speed control \cite{Hanson:2019}. These and other typical WIP algorithms are designed with the goal of making virtual walking more natural and closer to real walking, based on actual gait analysis. In general, the maximum controllable speed of WIP is approximately 2.5 m/s, and WIP faces difficulties in achieving a higher speed, that is, a ``running" speed. This is because the faster the virtual speed, the faster stepping must be applied by the users, thereby increasing not only their fatigue but also their risk of injury. Faster stepping also increases the shaking of the head-mounted display (HMD). Langbehn et al. introduced another control factor to improve WIP speed. They proposed leaning-amplified-speed WIP (LAS-WIP) that amplifies virtual walking speeds by the torso leaning angle \cite{langbehn2015evaluation}. However, they did not evaluate the virtual speed range that users can stably control with LAS-WIP.

\subsection{Output of WIP}

The main output of WIP is virtual movement. However, we have already discussed simple output movements calculated from the algorithm of WIP in Section 2.1. Therefore, in this section, we focus more on other outputs of WIP such as for simulating VEs.

Major studies on other outputs of the WIP investigate haptic feedback. These studies can be classified into two main groups. One group of studies uses shoe-type devices. Magana and Velazquez \cite{Magana:2008} confirmed the potential of haptic presentation on the sole of the foot using a shoe device equipped with a 16-point array of vibrotactile actuators. Moreover, Papetti et al. \cite{Papetti:2010} utilized miniature loudspeakers and broadband vibrotactile transducers, and Son et al. \cite{Son:2018} used magnetorheological fluid to represent virtual terrain textures. The second group uses floor-implemented devices, which include floor-type haptic displays that can present vibrotactile feedback \cite{Visell:2008,Law:2008} or tilt \cite{Bouguila:2003, Ishikawa:2018}. Most of these devices focus on the representation of specific terrain textures, and there is a trade-off between the range of haptic representations and the bulkiness of the device. 

In contrast, few methods for audio and visual output of WIP have been explored. Ishikawa et al. \cite{Ishikawa:2018} proposed a method to improve the perceived virtual slope gradient by tilting both the virtual camera (vision) and the treadmill (haptics); however, they did not find positive results in their experiment.

\subsubsection{Output of real walking locomotion}
In contrast to the few studies on the output of WIP, research on the output of real walking locomotion has been actively conducted; which could be used for WIP techniques. For example, a study on auditory feedback found that by adjusting the volume and timing of footsteps to the walking motion, it is possible to improve the presence and awareness of the user’s posture in VR \cite{Hoppe:2019}. 
%
In addition, there are studies on visual feedback that aim to represent a slope in VR by adjusting the physical load with a change in the ratio of the amount of physical movement to that of virtual movement \cite{Matsumoto:2017, Shimamura:2019}. 
%
Finally, Marchal et al. \cite{Marchal:2010} investigated the effect of visual parameters such as camera height, tilt, and speed on the simulation of bumps and holes located on the virtual ground. 
Although some of these methods could be implemented with WIP, their effects might differ between real walking and in-situ stepping.


\section{Investigating the Control Parameter: SHeF-WIP}

In this section, we propose a novel WIP control method called step height and frequency WIP (SHeF-WIP). We describe an experiment that examines the effects of SHeF control on users' WIP control strategies and performance.

\subsection{Technical Contribution}
The SHeF-WIP method relies on step height and frequency. We hypothesized that users can decrease their step frequency by raising their foot and therefore achieve a higher virtual speed with slower stepping motion, leading to more stability and safety at high speeds. Furthermore, we hypothesized that the step height parameter improves the naturalness and realism of virtual walking because we control not only the step frequency but also the step length in real walking. Therefore, our intention was to introduce a second control strategy of WIP in the standard WIP technique (GUD-WIP), enabling a richer UX and more varied strategies.

\subsubsection{Model}


SHeF is based on the GUD-WIP algorithm and introduces a step height factor. Equation (\ref{eq:gud}) shows the GUD-WIP algorithm, and Equation (\ref{eq:shef}) shows our SHeF-WIP algorithm.

\begin{equation}
|v_g| = \left(\frac{f}{1.57} \times \frac{H}{1.72}\right)^2 ~~~~~ [m/s]
\label{eq:gud}
\end{equation}

\begin{equation}
|v_s| = |v_g| \times \frac{sh}{Standard Step Height} ~~~~~ [m/s]
\label{eq:shef}
\end{equation}
Here, $|v_g|$ is the virtual speed of GUD-WIP, $|v_s|$ is the virtual speed of SHeF-WIP, $f$ is the step frequency computed in real time, $H$ is the height of the user, and $sh$ is the step height computed in real time.

The GUD-WIP algorithm refers to the relationship between the real walking speed, step frequency, and subject height, as proposed by Dean \cite{Dean:1965} and based on the analysis of human walking data. As shown in Equation (\ref{eq:gud}), the virtual speed of the GUD-WIP algorithm is based on the step frequency. This means that the higher the step frequency, the higher the virtual speed. The method of calculating the step frequency from the stepping gestures illustrates the originality and efficacy of GUD-WIP. It recognizes the user's stepping gestures as one of three states: grounded, ascending, or descending for each frame. By separating one-step gestures for the three states, GUD-WIP can estimate the step frequency with only a fraction of the completed step. Our SHeF-WIP approach adopts the same approach for computation of the step frequency. 

The SHeF-WIP algorithm adds a step height factor to the GUD-WIP algorithm (Equation (\ref{eq:shef})). Here, we determined the standard step height as 0.1 m in our pilot test: four participants walked through a straight and flat pass in a VR environment with GUD-WIP following a target moving at one of five constant speeds (0.5, 1.0, 1.5, 2.0, and 2.5 m/s). The average step height during the task at all speed levels was 0.094 ± 0.025 m. 
Then, we determined to use 0.1 m a standard step height during the WIP, allowing the relationship between step height and amplification of the virtual speed to be easily understood. From Equation \ref{eq:shef}, the virtual speed follows the GUD-WIP algorithm while users maintain a 0.1 m step height and is increased/decreased depending on a higher/lower step height. This means that the virtual speed of SHeF-WIP is two or three times the speed of GUD-WIP when the user's step height is 0.2 or 0.3 m, respectively.

Some other studies propose a WIP technique that uses step height as a parameter \cite{Nilsson:2013, Bruno:2013}. These methods use step height as a substitution for step length in real walking for more natural WIP experiences. On the other hand, SHeF-WIP introduces step height as a second control parameter mostly for a wider range of attainable virtual speeds. Therefore, the step height parameter in SHeF-WIP does not refer to real walking, unlike the other studies.

\subsection{Evaluation}
We conducted an experiment to evaluate SHeF-WIP by comparing it with the most common and strongly related WIP technique, GUD-WIP. In this experiment, we focused only on the evaluation of the control method, and not on the elastic or speed gain. Participants were asked to walk in VR at a constant speed while chasing a moving target in front of them. Subsequently, they answered questions about the walking experience. The hypotheses of the experiment are as follows: 

\begin{itemize}[\IEEEsetlabelwidth{\textrm{[H1-3]}}]
    \item [\textrm{[H1-1]}] Users can utilize the step height control.
    \item [\textrm{[H1-2]}] SHeF-WIP can achieve a wider range of controllable speeds with more stability than GUD-WIP.
    \item [\textrm{[H1-3]}] SHeF-WIP provides better UX such as realism, cyber sickness, and usability than GUD-WIP.
\end{itemize}

We have \textbf{H1-1} because we control both step frequency and step length in real walking, and we hypothesized that the controls of step height and length are similar. Furthermore, we have \textbf{H1-2} because the SHeF-WIP enables x times faster/slower output speed of GUD-WIP by step height parameter, and SHeF-WIP requires a lower step frequency (thus, slower and more stable gesture) if users lift their foot higher. Moreover, we have \textbf{H1-2} because we hypothesized that step frequency and step height control is closer to the real walking control than just step frequency control.

\subsubsection{Conditions}
The tested conditions were the WIP control method (GUD-WIP and SHeF-WIP), and the target walking speed was the speed in the six speed conditions of 0.5, 1.0, 1.5, 2.5, 3.5, and 4.5 m/s. The lower and middle speed levels of 0.5 to 1.5 m/s were used to see whether SHeF-WIP allows user control at a lower speed with similar stability as conventional WIP. Under these conditions, we wanted to confirm that SHeF-WIP not only enabled higher speeds but was also suitable for a wider range of speeds. A speed of 2.5 m/s was nearly the speed limit of the previous WIP schemes \cite{Wendt:2010,Bruno:2013}. The higher speeds of 3.5 and 4.5 m/s were used to determine the maximum speed of SHeF-WIP.

\subsubsection{Apparatus}

\begin{figure}[t!]
 \centering
 \includegraphics[width=50mm]{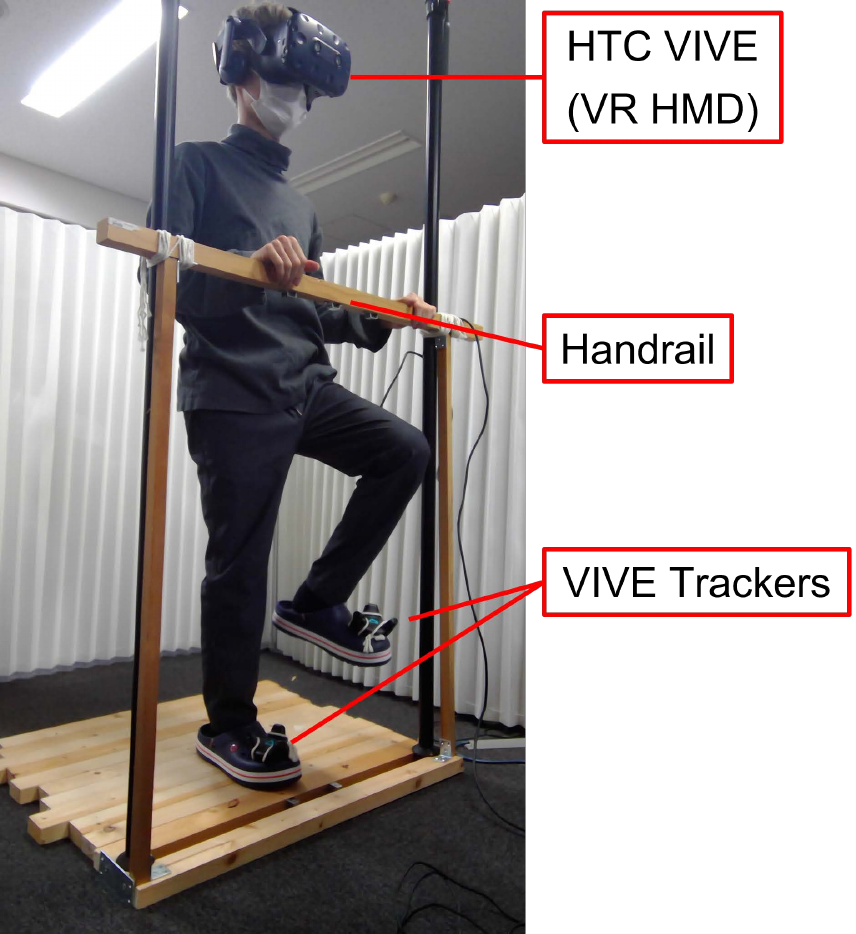}
 \caption{Physical apparatus of the system.}
 \label{Physical Apparatus}
\end{figure}

Fig.\ref{Physical Apparatus} shows the experimental apparatus. The participants put on a VR HMD (HTC VIVE Pro) and shoes with mounted VIVE trackers. Note that all participants used the same shoes. Four base stations were used: two were above the head for tracking the HMD, and the other two were placed on a table to track the foot trackers. The participants were asked to hold the handrail placed in front of them during the task to avoid walking outside of the tracked area of the base station and to avoid falling. 

\subsubsection{Task}

\begin{figure}[t]
 \centering
 \includegraphics[width=\columnwidth]{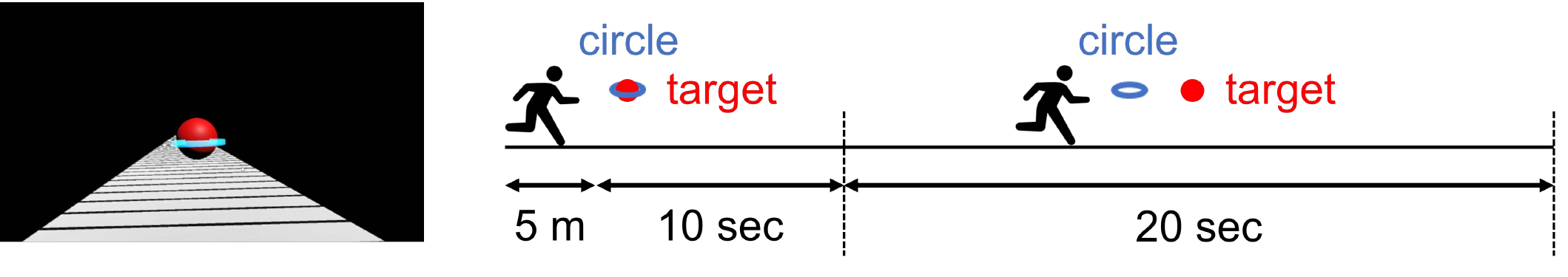}
 \caption{Participant's view (left) and an overview of the experimental scene (right) in VR. The target sphere and the circle moved at the same speed as the participants for the first 5 m and during a period of 10 s. In the following 20 s, the target sphere independently moved at one of the speed levels.}
 \label{XP1 VE}
\end{figure}

Fig.\ref{XP1 VE} shows the experimental scene in the VR environment. The participants stood on a straight and flat pass, facing a red sphere and a blue circle in front of them. The black lines were added to the path to make the participants feel the speed of movement in the simple VE. The red sphere was a target object moving forward at a constant speed, and the blue circle was an index of the participant’s virtual speed, that is, the blue circle continuously maintained a distance of 1 m from the participant. The task was to chase the target at a constant speed. Participants followed the red sphere and tried to keep it within the blue circle. This means that participants had to walk at the same speed as that of the red sphere while maintaining a constant distance from it. Both the radius of the red sphere and the inner radius of the blue circle are 0.25 m, so the sphere can fit within the circle.

The first 5 m and the following 10 s are the preparation stage in which the participants were asked to walk forward freely. During this stage, the red sphere and blue circle moved at the same speed as the participant.
After the preparation stage, a 3 s countdown was visually presented in front of them, and then the red sphere moved independently at one of the following speed levels: 0.5, 1.0, 1.5, 2.0, 3.5, or 4.5 m/s. The participants were asked to follow the red sphere and try to keep the red sphere within the blue circle. The red sphere stopped 20 s after the end of the countdown, which was the end of the task. The participants could see that a dummy translucent blue sphere repeatedly moved from the start point to the 5 m point at the target speed under the current condition. Therefore, participants knew the target speed in advance.

\subsubsection{Collected data}

The collected data is as follows:
\begin{itemize}
  \item The foot tracker positions in VR to calculate the stepping frequency and maximum height for each step.
  \item The distance between the center of the target and that of the blue circle.
  \item The walking speed to obtain the average walking speed and standard deviation of the walking speed.
  \item Questions with 6-point scale.
\end{itemize}

The sampling rate of each data was 90 hz.
Note that the standard deviation does not indicate that of the average walking speed. The standard deviation of walking speed was calculated from the logged walking speeds during the task for each individual participant. It is used to gain insight into how stably the participants could maintain a constant speed.

The participants were asked to answer questions on a 6-point scale from 1 (strongly disagree) to 6 (strongly agree) after each different WIP condition. Considering the mental load of participants and experiment time, we used a shorter custom questionnaire, rather than using a combination of existing questionnaires, because this would require the participants to answer a large number of questions. The questions were 1) I could control the virtual speed as intended (controllability), 2) I felt a sensation of walking (sense of walking), 3) I felt that the walking experience in the VE was similar to that in real life (realism of walking), 4) I became tired (tiredness), 5) I felt sick (cybersickness), and 6) I like the system (preference). After all trials, the participants were asked to answer their preferred WIP technique and their reasons for that preference.

\subsubsection{Procedure}
Twelve participants (10 males and 2 females, 11 in their 20s and one in her 30s) participated in the experiment. The mean height of the participants was $171.2 \pm 12.7 (cm)$ First, the experimental objectives, methods, and procedures were explained to them. Then, the participants completed the consent form and attribute sheet. In the attribute sheet, they provided their age, gender, and height. Next, the participants put on shoes with mounted trackers and practiced the stepping gesture. During the explanation of the WIP, the experimenter simply stated how WIP works, including the explanation of step frequency and height modulation. Moreover, the participants were requested to step freely to avoid biasing their stepping gestures. The participants then put on a VR HMD and practiced the experiment task three times each with GUD-WIP and SHeF-WIP at a speed of 1.5 m/s. After the practice, the experiment was conducted according to the following procedure: First, one WIP condition was randomly chosen by the experimenter, and the order was counterbalanced between the participants. The participants then conducted the task described in Section 3.2.3 for all six speed conditions in the first WIP condition. The order of the speed conditions was randomized. After the six trials, the participants took off the HMD and their shoes and completed the questionnaire online using their smartphones. After a short break, the participants repeated the same tasks under the other WIP condition and completed the same questionnaire. After all tasks were concluded, the participants provided their preference regarding the WIP method and their reasons for the preference. The participants were remunerated with an Amazon gift card of approximately 10 USD for their participation.

\subsection{Results}
\subsubsection{Stepping behavior} 

\begin{figure*}[t]
 \centering
 \includegraphics[width=\textwidth]{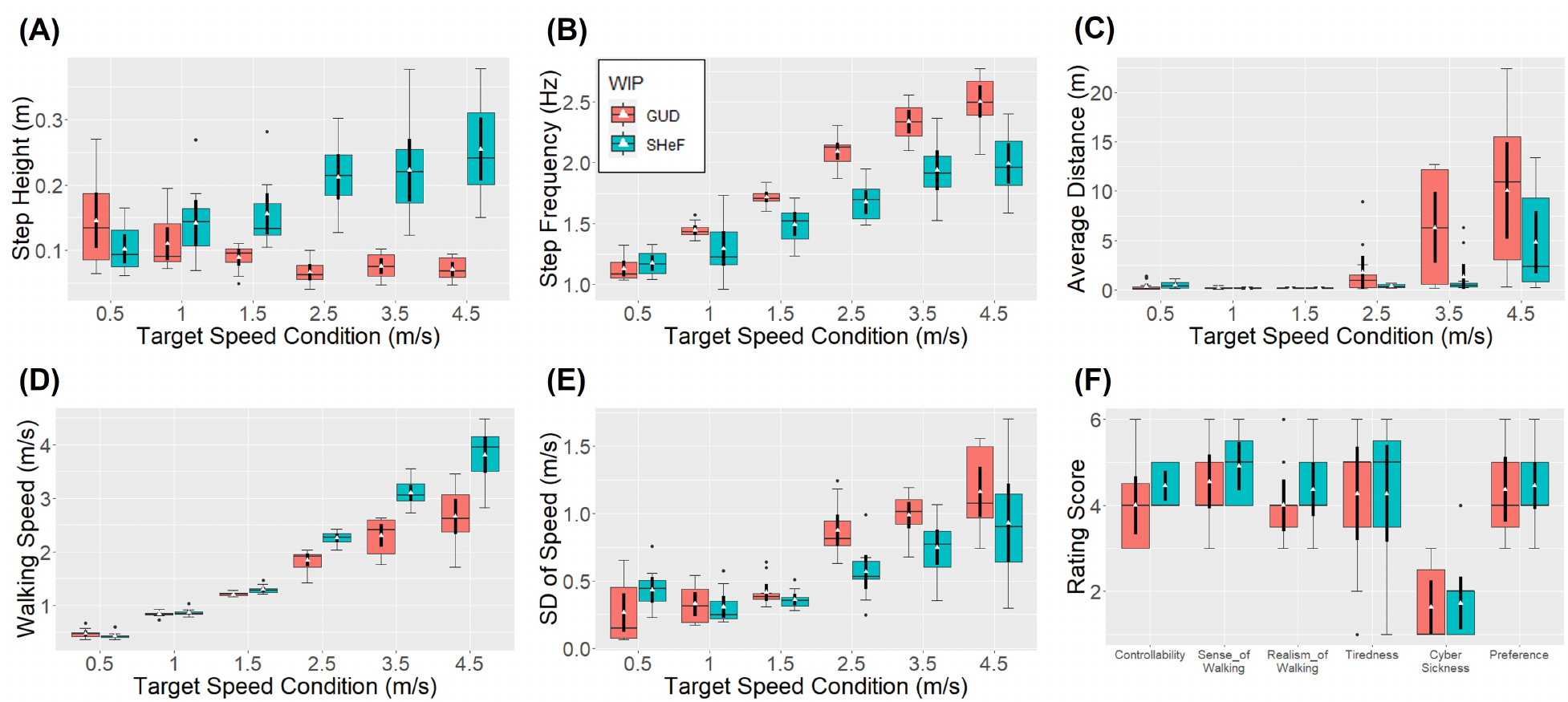}
 \caption{Results of Experiment 1. The white triangle indicates the mean, and the bold line shows the 95\% confidence interval. The error bar indicates the theoretical minimum or maximum value. Regarding a box plot, the middle horizontal lines show the medians, the colored boxes show the upper and lower quartiles, and the dots outside of the error bar show outliers. From (A) to (E), the horizontal axis represents the target speed conditions, and in (F), it represents each questionnaire item. The vertical axes indicate (A) the average step height during the chasing task; (B) the average step frequency during the chasing task; (C) the average distance from the center of the red sphere to that of the blue circle during the chasing task; (D) the average walking speed during the chasing task; (E) the standard deviations of the individual walking speeds during the chasing task; and (F)the rating scores for the questions.}
 \label{XP1_ALL}
\end{figure*}


Figs.\ref{XP1_ALL} (A) and (B) show the results of the average step height and average step frequency during the chasing task. A two-way ANOVA (\textit{WIP control} v.s. \textit{Target Speed}) was used to analyze the results. When the normality assumption was violated (Shapiro-Wilk test), Aligned Rank Transform (ART) was used for the data. When needed, Holm correction for p-value was used. Regarding the results of the step height, a significant main effect was observed in both the WIP factor ($F(1, 11)=162, p<0.001, \eta^2=0.48$) and the speed factor ($F(5, 55)=3.63, p=0.004, \eta^2=0.10$), and a significant interaction effect was found ($F(5, 55)=24.0, p<0.001, \eta^2=0.41$). Post-hoc tests (Wilcoxon signed rank test) showed that for speed conditions under the GUD-WIP condition, the average step height at 2.5 m/s was lower than that at 0.5 ($p=0.034$) and 1.0 ($p=0.034$), that at 4.5 m/s was lower than that at 0.5 m/s ($p=0.007$), and that at 3.5 m/s was lower than that at 1.0 m/s ($p=0.041$). In contrast, for the SHeF-WIP condition, the post-hoc tests showed that the following pairs of speed conditions were statistically different: 0.5-1.0 ($p=0.009$), 0.5-1.5 ($p=0.012$), 0.5-2.5 ($p=0.007$), 0.5-3.5 ($p=0.007$), 0.5-4.5 ($p=0.007$), 1.0-2.5 ($p=0.007$), 1.0-3.5 ($p=0.007$), 1.0-4.5 ($p=0.007$), 1.5-2.5 ($p=0.012$), 1.5-3.5 ($p=0.012$), 1.5-4.5 ($p=0.012$). Moreover, the test of no correlation was conducted between the height of participants and step height under one of the target speed conditions and one of the WIP methods. Consequently, no statistical differences were observed.

Regarding the results of the step frequency, a significant main effect was observed in both the WIP factor ($F(1, 11)=106, p<0.001, \eta^2=0.28$) and speed factor ($F(5, 55)=193, p<0.001, \eta^2=0.55$), and a significant interaction effect was also found ($F(5, 55)=10.4, p<0.001, \eta^2=0.18$). The post-hoc tests for multiple comparisons for the speed condition showed that all pairs of speed conditions were statistically different (3.5-4.5: $p=0.018$, other pairs: $p<0.001$). Furthermore, a post-hoc test for the simple effect of the WIP factor at each speed revealed that the average step frequency of GUD-WIP was statistically higher than that of SHeF-WIP at 1.5 ($p<0.001$), 2.5 ($p<0.001$), 3.5 ($p=0.002$), and 4.5 ($p<0.001$) m/s conditions.

\subsubsection{Participant following of target}

Fig.\ref{XP1_ALL} (C) shows the average distance from the center of the red sphere to that of the blue circle during the chasing task. Lower values of the average distance mean that participants were able to follow the target more accurately. A two-factor ANOVA was conducted. A significant main effect was observed in both the WIP factor ($F(1, 11)=54.0, p<0.001, \eta^2=0.33$) and speed factor ($F(5, 55)=15.5, p<0.001, \eta^2=0.40$). A significant interaction effect was also found ($F(5, 55)=8.65, p<0.001, \eta^2=0.27$). A post-hoc test for the effect of the WIP factor at each speed revealed that the average distance in SHeF-WIP was significantly lower than that in GUD-WIP under the speed conditions of 3.5 m/s ($p=0.012$). 

\subsubsection{Walking speed and stability of control}

Fig.\ref{XP1_ALL} (D) presents the average walking speed during the chasing task. Fig.\ref{XP1_ALL} (E) shows the standard deviation of individual walking speeds during the task, which indicates the stability of the participant's walking speed. A two-way ANOVA was performed for both results.

Regarding the results of the walking speed, a significant main effect was observed for both the WIP factor ($F(1, 11)=135, p<0.001, \eta^2=0.25$) and speed factor ($F(5, 55)=561, p<0.001, \eta^2=0.48$). In addition, a significant interaction effect was found ($F(5, 55)=33.2, p<0.001, \eta^2=0.27$). The results of the post-hoc tests suggested that the walking speed of SHeF-WIP was significantly faster than that of GUD-WIP at speeds of 1.5 ($p=0.002$), 2.5 ($p<0.001$), 3.5 ($p<0.001$), and 4.5 ($p<0.001$) m/s. This result shows that SHeF-WIP enables users to achieve a walking speed closer to the target speed when moving at moderate and high speeds.

Regarding the results of the individual standard deviation of the walking speed, a significant main effect was observed for the WIP factor ($F(1, 11)=13.9, p<0.001, \eta^2=0.10$) and speed factor ($F(5, 55)=66.1, p<0.001, \eta^2=0.73$), and a significant interaction effect was found ($F(5, 55)=5.37, p<0.001, \eta^2=0.17$). The post-hoc test indicated that the individual standard deviation of the walking speed during the task of SHeF-WIP was significantly smaller than that of GUD-WIP in the speed conditions of 2.5 ($p<0.001$) and 3.5 ($p=0.009$) m/s. These results show that SHeF-WIP enables participants to control their walking speed more precisely at high speeds.

\subsubsection{Questionnaires and preference}

The questionnaire responses of one participant could not be obtained because of a network issue during the experiment; therefore, the following analysis was conducted with the results of 11 participants. An exact Wilcoxon signed-rank test was applied to all questionnaires, and no significant differences were found. In the final question about the preference of WIP, seven of the participants preferred SHeF-WIP and five preferred GUD-WIP. The reasons reported for this are as follows (the number after each comment indicates the number of participants who mentioned it): SHeF-WIP made it easier to reach a higher speed with less effort/tiredness (4). SHeF-WIP was more natural or closer to real walking (4). It was easier to control the speed with SHeF-WIP (2), or easier with GUD-WIP (5). Focusing on both the step frequency and step height was difficult, compared with considering the frequency only (3). The frequency strategy was easier to control than step height strategy (1).

\subsection{Discussion}

\subsubsection{Can users utilize the step height control?}
\textbf{H1-1} was supported. As shown by the results of the average step height and frequency during the chasing task (Figs.\ref{XP1_ALL} (A) and (B)), participants changed their stepping strategy for SHeF-WIP depending on the required virtual speed. Fig.\ref{XP1_ALL}(A) indicates that the higher the desired speed, the higher the participants raised their feet, whereas the average step height was kept constant or lower in GUD-WIP. In addition, Fig.\ref{XP1_ALL}(B) shows that the average step frequency in SHeF-WIP was kept lower than that of GUD-WIP. These results indicate that users can utilize both the step height and frequency strategies for virtual speed control. Moreover, the results of the no-correlation test between participants' height and step height indicate that the height has a small effect on the step height control of SHeF-WIP.

\subsubsection{Can SHeF-WIP achieve a wider range of speed with more stability than GUD-WIP?} 
Analysis of the results for the average distance (Fig.\ref{XP1_ALL}(C)), the average walking speed (Fig.\ref{XP1_ALL}(D)), and the individual standard deviations of the speed (Fig.\ref{XP1_ALL} (E)) support \textbf{H1-2}. As shown in Fig.\ref{XP1_ALL} (C), the participants successfully achieved a speed of 3.5 m/s with SHeF-WIP, whereas with GUD-WIP, the participants started to face difficulties in catching up with the target at 2.5 m/s. A similar conclusion can be obtained from Fig.\ref{XP1_ALL} (E): the speed control was significantly more stable with SHeF-WIP than with GUD-WIP in the speed conditions of 2.5 and 3.5 m/s. Considering Figs.\ref{XP1_ALL} (C) and (E), we can roughly estimate that the ceiling of virtual speed that users can control with stability is the speed in the 2.5 m/s condition for GUD-WIP and the speed in the 3.5 m/s condition for SHeF-WIP. These speeds were 1.84 $\pm$ 0.18 m/s for GUD-WIP and 3.10 $\pm$ 0.23 m/s for SHeF-WIP (Fig.\ref{XP1_ALL} (D)). This estimated maximum speed of GUD-WIP is similar to the result of Bruno et al. \cite{Bruno:2013}, where the speed was 1.93 $\pm$ 0.23 m/s; however, they did not consider stability. In addition, considering that the average step frequency with SHeF-WIP was significantly lower at high speeds (Fig.\ref{XP1_ALL}(B)) and that SHeF-WIP enabled more stable control of lower speeds than GUD-WIP (Fig.\ref{XP1_ALL} (E) ), we can conclude that SHeF-WIP can achieve a wider range of virtual speeds (about +80\%) with more stability and slower motion because of lower stepping frequency. We suggest that this results in a richer experience and probably less fatigue and a lower risk of injury than GUD-WIP.


\subsubsection{Does SHeF-WIP provide better UX than GUD-WIP?}
\textbf{H1-3} was not supported by the results of the questionnaires. 
Note that our experimental task and VE are designed mainly for establishing how precisely the participants could adjust the speed using the two WIP techniques. However, they might not be optimal for assessing the realism of the techniques. Still, both GUD-WIP and SHeF-WIP achieved relatively good scores on each question, excluding ``tiredness.'' Therefore, we would conclude that SHeF-WIP can achieve higher speed than GUD-WIP, while ensuring controllability, sense of walking, and sense of realism, without improving the occurrence of cybersickness. However, we need to be careful about the results of the questionnaires due to the small sample size.

Based on these results, the next section will investigate the effects of step height control and elastic input on the performance and UX of the WIP.

\section{Investigating the Input Parameter: Elastic Input for WIP}

In this section, we propose a novel passive and elastic input system for WIP. We describe an experiment on the effects of the SHeF control and the elastic input system on the WIP performance and UX. 

\subsection{Technical Contribution}

\begin{figure}[b]
 \centering
 \includegraphics[width=\columnwidth]{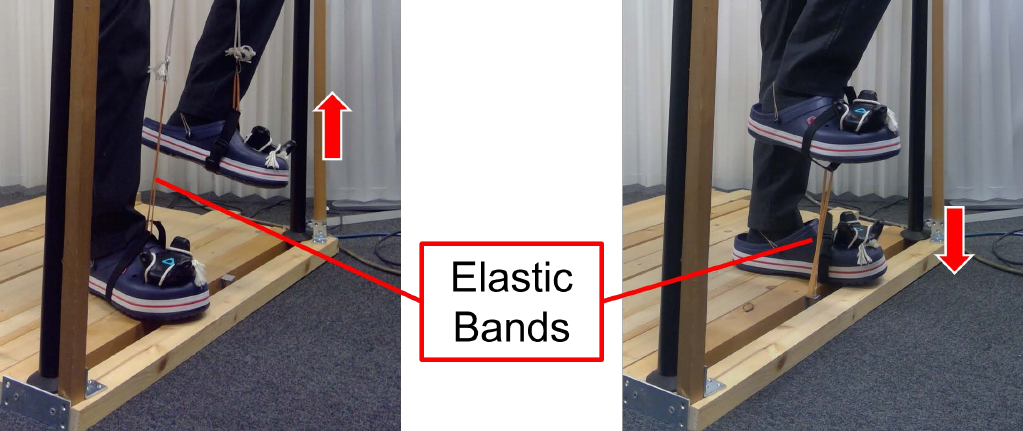}
 \caption{Elastic input system. The user's feet are pulled upward (left figure) or downward (right figure) by elastic bands.}
 \label{Elastic Bands Apparatus}
\end{figure}

We introduced a vertical elastic input to the SHeF-WIP as a novel input parameter. The elastic input provides users with different levels of elastic force, depending on their step height. Therefore, we hypothesized that the elastic input helps users obtain a better sense of their step height, which leads to better control of SHeF-WIP. To the best of our knowledge, no research has investigated the input of WIP using stepping gestures to improve the performance of WIP, such as how stably users can control the virtual speed. 

Fig.\ref{Elastic Bands Apparatus} shows the setup of the elastic input. The left figure indicates the elastic force presentation in the upward direction and the right figure indicates the downward direction. The elastic force and the directions are easily adjustable by adding or removing elastic bands. In the downward mode, the elastic bands are attached to the hook on the ground and shoes. In the upward mode, they are attached to the hook on the rod and the shoes. The distance from the hook to the shoes on the ground was 39 cm and it was sufficient height as there was always tension on the rubber band while participants perform a stepping gesture. This distance was decided considering both the step height results of Experiment 1 and the elastic force levels described in Section 4.2.3. Elastic bands can be easily attached to and detached from the shoes by using a strap on the shoes.

We used elastic bands of Cemedine XA-129 Rubber Band No. 16 (Cemedain Co., Ltd). This rubber band is easily available and stretchable, and therefore it is simple to adjust the force by changing the number of bands. The folded length of the rubber band was 6 cm. We measured the relationship between its extension and force using three new and three used rubber bands. The used rubber bands were the ones after repeatedly being stretched for 1 minute at almost maximum extension (around 40 cm). To measure the length/force relationship, a spring scale was used. The results showed that the relationship is relatively stable in the extension range of under 25 cm regardless of the newness. The approximate curve of the relationship is $f=0.011*e+0.085 (0 \leq e \leq 25), R^2=0.96$, where f indicates the force (kg) and e indicates the extension (cm). Based on this relationship, we simulated the required force presented to the user during WIP. 

\subsection{Evaluation}
We conducted an experiment to evaluate the performance and UX of the WIP with elastic input and to determine the appropriate elastic force and its direction for WIP. In this experiment, the SHeF-WIP method was used considering the synergy between the step height control for the WIP and elastic input. The experimental hypotheses are as follows:

\begin{itemize}[\IEEEsetlabelwidth{\textrm{[H2-2]}}]
    \item [\textrm{[H2-1]}] Passive elastic force enhances the stability of WIP speed control.
    \item [\textrm{[H2-2]}] Not too strong passive elastic force enhances the UX such as realism and usability.
\end{itemize}

We have \textbf{H2-1} and \textbf{H2-2} because we hypothesized that the elastic force gives users a better sense of their step height, allowing users to have a better command of WIP control. Here, we did not have hypotheses related to whether upward or downward feedback might be the most beneficial because we found the benefit of elastic input in a point that it can provide a relationship between haptic force intensity and step height. Therefore, we had the condition of elastic direction for an exploratory objective.

\subsubsection{Conditions}
In the experiment, we compared two directions of elastic input (upward/downward) and three different elastic forces for each direction condition (weak/middle/strong). In addition, we prepared a normal walking condition without an elastic input. Therefore, the total number of experimental conditions was seven. We labeled these conditions as; DS, downward strong; DM, downward middle; DW, downward weak; N, normal; UW, upward weak; UM, upward middle; and US, upward strong. The elastic strength values were approximately 1, 2, and 3 kg at a 15.6 cm step height in the downward condition (5, 10, and 15 rubber bands for each condition) and approximately 1, 3, and 5 kg at a step height of 0 cm in the upward condition (3, 9 and 15 rubber bands for each condition). The 15.6 cm height was the average step height during WIP at a speed of 1.5 m/s according to the results of Experiment 1 (Fig.\ref{XP1_ALL}(A)). These values were chosen as weak, middle, and strong elastic forces based on a pilot experiment.

\subsubsection{Apparatus}
The physical apparatus is shown in Figs.\ref{Physical Apparatus} and \ref{Elastic Bands Apparatus}. We replaced the elastic bands with new ones for each participant to avoid their deterioration. During the experiment, participants put on a VR HMD and shoes, held the rod in front of them, and performed marching gestures to walk forward in VR. The VE was the same as that of Experiment 1.

\subsubsection{Task}
The task was almost the same as that in Experiment 1. In this experiment, the target sphere's speed was fixed at 1.5 m/s in the main chasing part for all conditions. Moreover, as a training task, participants were asked to perform the main task explained in Section 3.2.2, where the target's speed dynamically changed from 1.0 m/s to 2.0 m/s. This training task was conducted before every first trial of each elastic condition for the participants to adapt to the elastic force.

\subsubsection{Collected data}
The same behavioral data as in Experiment 1 were recorded. In addition, participants were asked to answer questions on a 6-point scale from 1 (strongly disagree) to 6 (strongly agree) after each trial. Considering the mental load of participants and experiment time, we used a shorter custom questionnaire, rather than using a combination of existing questionnaires, because this would require the participants to answer a large number of questions. The items were 1) I felt that the walking experience in VR was similar to that in real life (realism of walking), 2) I felt a sensation of walking (sensation of walking), 3) I got tired (fatigue), and 4) the questions from system usability scale (SUS) \cite{Brooke:1996}. In addition, we asked the participants about their preference of the strength (weak/middle/strength) for each direction condition (upward/downward) and the reason for their choice.

\subsubsection{Procedure}
Twenty participants (19 males and 1 female in their 20s) participated in the experiment. The mean height of the participants was $169.2 \pm 12.5 (cm)$. The participants put on the shoes, and we confirmed that elastic bands could be attached to or detached from the shoes. The participants practiced the stepping gesture without elastic input. Then, the participants put on a VR HMD and practiced virtual walking in VR without elastic input. After the practice, the main experiment was conducted according to the following procedure: First, one input condition was randomly chosen, and the elastic bands were attached to the shoes if necessary. Then, the participants conducted a training task to get used to the WIP under the input condition. After that, they performed the main task twice and then removed their HMD and answered the questions. If the condition was an upward or downward condition, they repeated the set of training tasks and main tasks, and completed the questionnaires for the other two elastic force levels in random order. They conducted this set of three trials (for the upward or downward condition) or one trial (for the normal condition) for each direction condition in random order. They repeated the above twice, and therefore, they performed the main task four times for each condition. At the end of the experiment, they answered the question regarding their preference and reasons. After the experiment, they were remunerated with an Amazon gift card of approximately 15 USD for their participation.

\begin{figure*}[t]
 \centering
 \includegraphics[width=\textwidth]{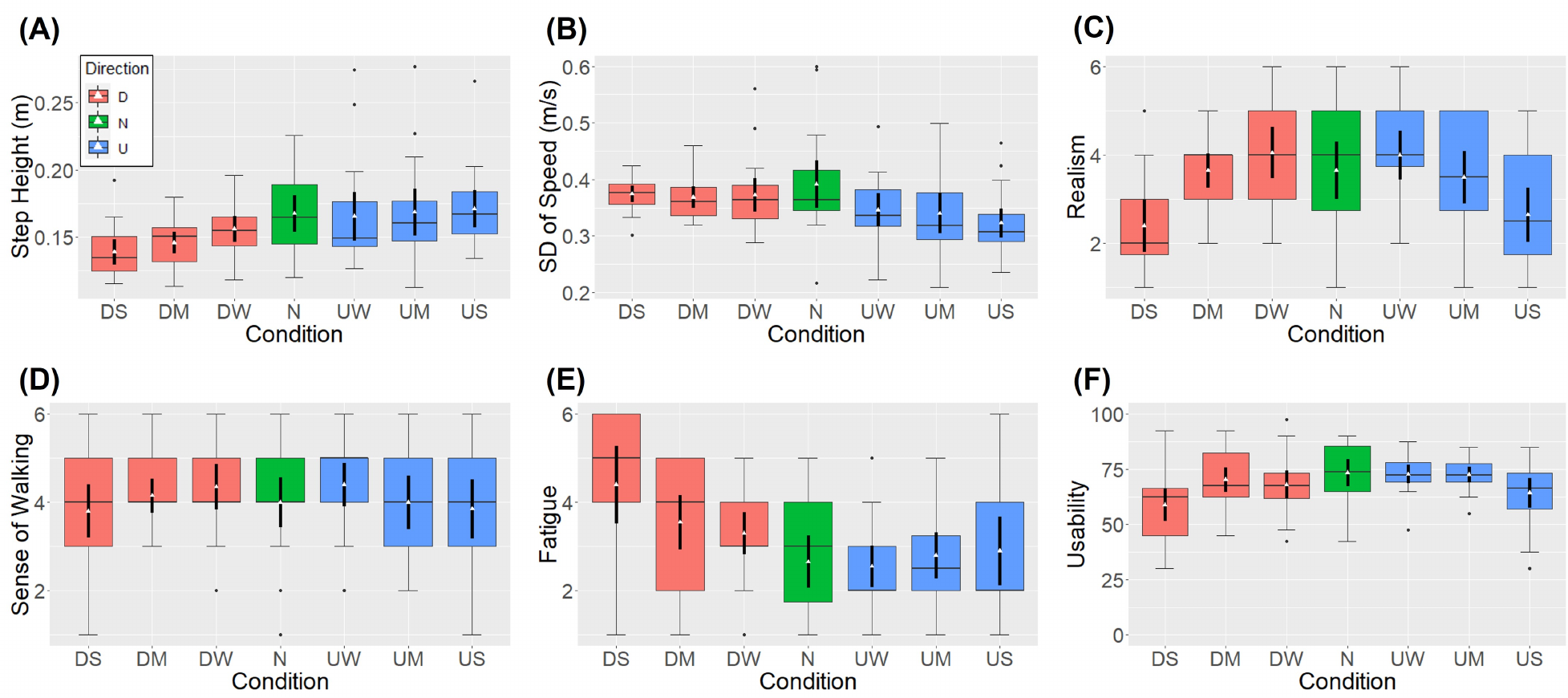}
 \caption{Results of Experiment 2. The white triangle indicates the mean, and the bold line shows the 95\% confidence interval. The error bar indicates the theoretical minimum or maximum value. Regarding a box plot, the middle horizontal lines show the medians, the colored boxes show the upper and lower quartiles, and the dots outside of the error bar show outliers. In all figures, the horizontal axis represents the elastic input conditions. The vertical axes indicate (A) the average step height during the chasing task; (B) the standard deviations of the individual walking speeds during the chasing task; (C) the rating score for the questionnaire item of realism; (D) the sense of walking; (E) the sense of fatigue; and (F) the SUS score for usability.}
 \label{ALL_XP2}
\end{figure*}

\subsection{Results}

\subsubsection{Stepping behavior}
Fig.\ref{ALL_XP2}(A) shows the average step height during the chasing task. The Wilcoxon signed rank test with Holm correction was conducted, and statistical differences were found between DS-DW ($p=0.015$), DS-N ($p=0.008$), DS-UW ($p=0.041$), DS-UM ($p=0.037$), DS-US ($p<0.001$), DM-UM ($p=0.024$), and DM-US ($p<0.001$) conditions.

\subsubsection{Performance of WIP}


The results of average distance (mean $\pm$ 2 standard deviation) were as follows: DS, $0.16 \pm 0.21$; DM, $0.16 \pm 0.16$; DW, $0.16 \pm 0.15$; N, $0.16 \pm 0.14$; US, $0.15 \pm 0.13$; UM, $0.17 \pm 0.15$; and US, $0.17 \pm 0.15$. The Wilcoxon singed rank test with Holm correction was conducted to determine the average distance from the center of the red sphere to that of the blue circle during the chasing task. No statistical differences were found between the conditions.

Fig.\ref{ALL_XP2}(B) shows the results of standard deviation of the walking speed for each participant, which indicates the stability with which the participants could keep a constant speed in each condition. {The Wilcoxon singed rank test with Holm correction} was conducted, and a significant difference was found between DS-US ($p=0.006$) and N-US ($p=0.002$). This result suggests that the upward-strong elastic input helps the participants maintain a constant 1.5 m/s walking speed better than the WIP with downward strong elastic input and without the elastic input.

\subsubsection{UX of WIP}

Figs.\ref{ALL_XP2}(C)–(E) show the responses to the questions. The Wilcoxon signed rank test with Holm correction was conducted for the questions about realism, sense of walking, and fatigue, and SUS scores calculated from the SUS questions. 

For realism, significant differences were found between DS-DM ($p=0.037$), DS-DW ($p=0.021$), and DS-UW ($p=0.041$). In addition, no significant differences were observed in the sense of walking. Furthermore, for fatigue, there were significant differences between DS-UW ($p=0.043$). Finally, for the usability, significant differences were found in the DS-DM ($p=0.042$), DS-UM ($p=0.041$), and UM-US ($p=0.040$).

Regarding the preference of the strength for each direction condition, the numbers of participants who preferred weak, middle, and strong were 16, 3, and 1, respectively, for the downward condition and 2, 12, and 6, respectively, for the upward condition. All comments from the participants who preferred downward weak condition were about fatigue, the realism of walking, or ease of WIP (usability). The four participants who chose middle or strong in the downward condition mentioned the stability of the WIP. In addition, most of the comments for upward conditions (17 out of 20) were that the upward elastic force supported their stepping gestures, and they felt comfortable with it. In addition, three participants who preferred middle strength in the upward condition mentioned that the elastic force improved the realism of walking.

\subsection{Discussion}
\subsubsection{Does appropriate passive elastic force enhance the stability of WIP speed control?}
\textbf{H2-1} was partially supported by the results of the  standard deviations of the individual walking speeds. First, the results of the average distance indicate that participants managed to catch up with a moving target in VR with all elastic conditions. According to the analysis of the standard deviation of the speed, the upward-strong elastic force improves the stability of the WIP control compared to the WIP without elastic input. However, this increment in the WIP control stability was not observed in the other conditions. In this experiment, we evaluated the conditions in the task of chasing a virtual object that moves at a constant speed. Here, maintaining a constant speed in the WIP means repeating the same stepping gesture periodically, which is a relatively easy task. Therefore, it is possible that there were no significant differences in the stability of walking speed between the other conditions. In future work, we will investigate the synergy between elastic input and step height control in a task that is more related to real use cases, such as the control of various walking speeds.


\subsubsection{Does appropriate passive elastic force enhance the UX of WIP?}
\textbf{H2-2} was not supported by our experimental results. The questionnaire results show that the UX and usability of WIP for a virtual flat road are already positive for WIP without elastic input. Our results suggest that middle or less passive elastic forces have small effects on the UX of WIP for a virtual flat road.

\subsubsection{Other implications}
As discussed above, the upward-strong elastic input helped users maintain a constant walking speed while decreasing usability. This indicates that there is a trade-off between performance and UX for the upward elastic input. In contrast, the downward-strong elastic input did not have a significant effect on the performance of WIP, while having a negative influence on UX. 
As the weak and middle strength elastic input did not have much influence on the stepping behavior, WIP performance, or UX of WIP on a virtual flat path, developers can use the elastic input to keep the users in the same position while playing VR content using WIP. Moreover, the weak and middle elastic inputs could be used to present haptic feedback in VR, which will be investigated in the next section. However, as a limitation of the current elastic input setup, the user cannot turn around with the system, and this issue should be solved for a natural WIP experience in future work.

Based on the above results, the next section investigates the effects of SHeF-WIP control, elastic input, and pseudo-haptic feedback on the virtual slope walking experience.



\section{Investigating the Output Parameter: Pseudo-Haptic Approach with WIP}


In this section, we first introduce a pseudo-haptic approach to the WIP to realize a virtual slope. We expect the synergy of passive elastic input and pseudo-haptics to improve the realism of walking. We describe an experiment investigating the effects of pseudo-haptic feedback for a virtual slope representation, with SHeF control and elastic input, on the realism of the VE and UX.

\subsection{Technical Contribution}
The pseudo-haptic technique is a method to modify force perception by mainly visual feedback \cite{Pusch:2011}. In this paper, the WIP output speed is modified by the speed gain, that is, the gain between the stepping gesture and output speed as shown in Equation (\ref{eq:pseudo-haptics}).
\begin{equation}
|v| = |v_s| \times g ~~~~~ [m/s]
\label{eq:pseudo-haptics}
\end{equation}
Here, $|v|$ is the output speed, $|v_s|$ is the calculated speed of the SHeF-WIP, and $g$ is the speed gain. When the speed gain is 1.0, the output speed is the same as that of the original SHeF-WIP. When the speed gain is less than 1.0, the output speed is slower than that of the SHeF-WIP and vice versa.

We hypothesized that a smaller speed gain would be suitable for VEs such as uphill or mud because the speed of the progress will be slower, thus increasing the user's perception of pseudo-load. In contrast, we expected that a larger speed gain could be applied to VEs such as downhill or icy roads because they make the user move faster, thus reducing the user's perception of pseudo-load.

Furthermore, pseudo-haptic feedback should have a synergy with the elastic input system \cite{Paljic:2004, paljic2004evaluation, Achibet:2014, Achibet:2015}, and pseudo-haptic feedback and SHeF-WIP control are also expected to affect each other. Pseudo-haptic feedback improves the realism of VR and it is suggested that the realism of VR would affect users' behavior \cite{Ogawa:2020}. Therefore, we hypothesized that the pseudo-haptic feedback would affect stepping gestures and if so, SHeF-WIP speeds would reflect behavioral changes better than step frequency control only.


\subsection{Evaluation}
We conducted an experiment to evaluate the pseudo-haptic approach and elastic input with SHeF-WIP to improve the realism of walking in VEs. In the first investigation, we examined its application to slopes, which are common in everyday space. The SHeF-WIP was used for the experiment. In addition, the experiment was conducted with the method of adjustment. Participants were asked to find an appropriate speed gain for a virtual downhill/uphill for each input condition, and then evaluate the walking experiences with the individualized speed gain for each condition. The hypotheses of this experiment are as follows.

\begin{itemize}[\IEEEsetlabelwidth{\textrm{[H3-2]}}]
    \item [\textrm{[H3-1]}] Pseudo-haptic approach with WIP can simulate a virtual slope.
    \item [\textrm{[H3-2]}] Elastic input with a pseudo-haptic approach increases the realism of a virtual slope.
\end{itemize}

Here, we consider that [H3-1] holds if participants can interpret the speed gain as pseudo-haptic load while walking on a slope, like previous studies \cite{Matsumoto:2017, Shimamura:2019}. Moreover, we have \textbf{H3-2} because several studies indicate that elastic force with pseudo-haptics improves the realism of VEs \cite{Paljic:2004, paljic2004evaluation, Achibet:2014, Achibet:2015}.

\subsubsection{Conditions}
Three input conditions were compared for each of the two slope conditions. The input conditions consisted of upward and downward elastic inputs of the middle strength used in Experiment 2 and the normal walking condition (without elastic input). In this experiment, the middle strength for elastic input was adopted after consideration of the preference and performance results from Experiment 2 and with the expectation of synergy of SHeF-WIP and pseudo-haptics for the realism of VEs. Furthermore, the two slope conditions were uphill and downhill.

\subsubsection{Apparatus}

\begin{figure}[t]
 \centering
 \includegraphics[width=\columnwidth]{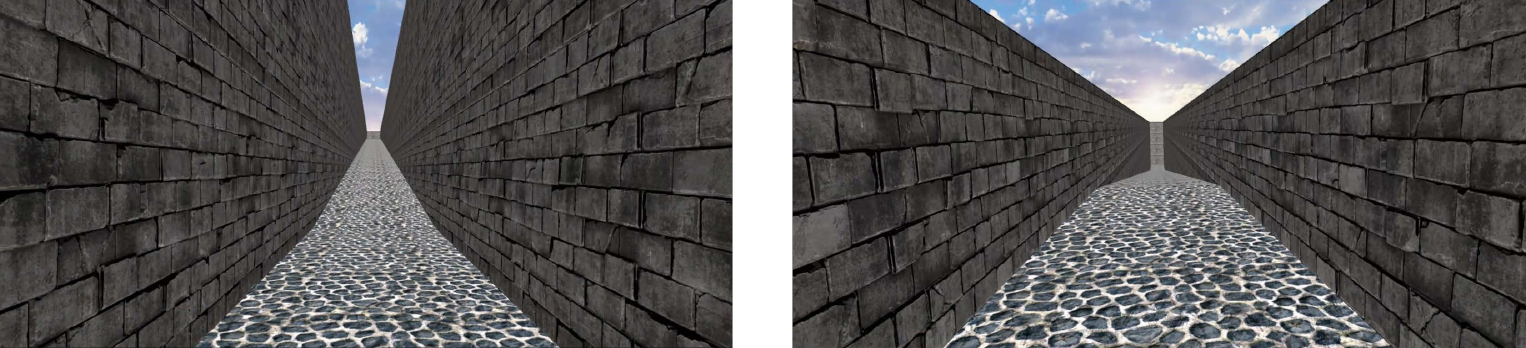}
 \caption{The VE for the experiment. The figures show the view of a participant at the uphill (left) and downhill (right) starting points.}
 \label{XP3 VE}
\end{figure}

The physical apparatus was the same as that used in Experiment 2; here, we focus on the explanation of VE. Fig.\ref{XP3 VE} shows the overview of the path in VR. The gradient of the slope is 5.71° (approximately 10\% gradient), which is a relatively steep slope in the physical world. Furthermore, the length of the slope was set to 75 m, and the participants were not expected to reach the end of the path during the task. The details of the task are explained in Section 5.2.2. In this experiment, the output of the virtual walking speed is 2.02 times faster than the pure calculated speed from the SHeF-WIP equation. This is because Nilsson et al. \cite{Nilsson:2014} confirmed that there ``exists a range of perceptually natural visual gains" and 2.02 $\pm$ 0.16 was the mean visual gain rated as natural for WIP with a HMD whose diagonal field of view (FOV) is 60 degree for each eye. In this experiment, we put more weight on the evaluation of participants' perception of virtual walking, while Experiments 1 and 2 focused on walking performance. Nilsson et al. \cite{Nilsson:2014} confirmed that less visual gain than 2.02 is enough with a HMD that has a wider FOV. However, our informal test with three participants confirmed that the virtual walking experience on a virtual flat road with SHeF-WIP is natural with the visual gain of 2.02 and with the VIVE pro. Moreover, this visual gain might change the values of our results but does not affect the general tendencies of the results and thus, does not change the discussion of the hypotheses or conclusion of this experiment. The speed gain was 1.0 on the first 10 m of the flat road, and the gain of the condition was applied instantly when the participants reached the edge of the slope.

\subsubsection{Task}
The experimental task was divided into two parts. The first part was conducted using the method of adjustment and participants adjusted speed gain to obtain a natural walking speed. In the method of adjustment, there were two series regarding the speed gain. One series started from a relatively small speed gain and the other started from a relatively large speed gain. In addition, different ranges of speed gains were used for uphill and downhill. The initial speed gains were 0.3 or 1.0 for uphill and 1.0 or 2.5 for downhill. The participants walked up/down a virtual hill for 5 s with an initial speed gain. Then, the participants were returned to the starting point and asked to increase or decrease speed gain. The initial speed gains were chosen from our preliminary experiment as a range such that the adjusted values were expected not to exceed them. In addition, the gain could be adjusted in steps of 0.07 for uphill and 0.15 for downhill to allow the participants to adjust the speed gain 10 times within the range. It should be noted that the participants could continue adjusting beyond the range if they wanted. The participants repeated this walking and adjustment trial until they felt that their walking experience was natural: the message that was presented to the participants in the VR was: { \it ``If you feel that the latest walking speed was appropriate, pull the trigger. If not, press the pad to walk up/down faster/slower."} In this first part, the individualized appropriate speed gain for each condition was obtained.


The second part was an evaluation of the virtual walking experience. In this part, the participants were asked to walk up/down a hill for 5 s with the individualized speed gain. They were then asked to answer questions evaluating their walking experiences, which are explained in Section 5.2.4.

\subsubsection{Collected data}
The adjusted speed gains were recorded to compute the appropriate speed gains for each condition. Moreover, the participants were asked to answer questions on a 6-point scale from 1 (strongly disagree) to 6 (strongly agree) in the evaluation task of the second part. Considering the mental load of participants and experiment time, we used a shorter custom questionnaire, rather than using a combination of existing questionnaires, because this would require the participants to answer a large number of questions. The items were 1) I felt as if I really went up/down a hill (realism of walking), 2) I felt the hill was steep (intensity of pseudo-haptics), and 3) In the VE, I had a sense of ``being there'' (sense of presence). It should be noted that in the second item, a score of 1 indicates ``not steep at all'' and 6 indicates ``extremely steep.''

\subsubsection{Procedure}
Eighteen participants (12 males and 6 females; 17 in their 20s and 1 in her 30s) participated in the experiment. The preparation procedure before the main task was the same as that used in Experiment 2. In the first part of the main task, the participants conducted two repetitions of each ascending and descending adjustment for each combination of the two slope conditions and three input conditions in random order. Therefore, the total number of trials was 24 (three input conditions $\times$ 2 slope conditions $\times$ 2 adjustment series $\times$ 2 repetitions). After the adjustment part, the participants took a break and then completed the evaluation task. In the second part (evaluation task), they performed six trials in total (three input conditions $\times$ 2 slope conditions). They were remunerated with an Amazon gift card of approximately 15 USD for their participation.

\subsection{Results}
\subsubsection{Adjusted speed gain}



Table \ref{XP3 table speed gain} shows the results of the adjusted speed gain. As explained in the procedure section, the speed gains were obtained 4 times for each condition from each participant (i.e., 2 repetitions for each ascending and descending series), and these speed gains were averaged. All average speed gains were below 1.0 for the downhill condition and above 1.0 for the uphill condition. We conducted a paired t-test with Bonferroni correction within each slope condition. 
Consequently, no statistical differences were found. Note that these gains were derived from the SHeF-WIP output $\times$ 2.02.

\begin{table}[h]
\renewcommand{\arraystretch}{1.3}
\centering
\caption{Appropriate speed gains for each condition, which were obtained in the adjustment task. The conditions in rows indicate slope conditions and the conditions in columns indicate elastic input conditions.}
    \begin{tabular}{|c|c|c|c|}  \hline
                 & Downward         & Normal        & Upward          \\ \hline
        Uphill & 0.75 $\pm$ 0.07 & 0.71 $\pm$ 0.07 & 0.74 $\pm$ 0.10 \\ \hline
        Downhill   & 1.54 $\pm$ 0.18 & 1.43 $\pm$ 0.25 & 1.49 $\pm$ 0.23 \\ \hline
    \end{tabular}
    \label{XP3 table speed gain}
\end{table}

\subsubsection{UX of WIP with a virtual slope}

\begin{figure*}[t]
 \centering
 \includegraphics[width=\textwidth]{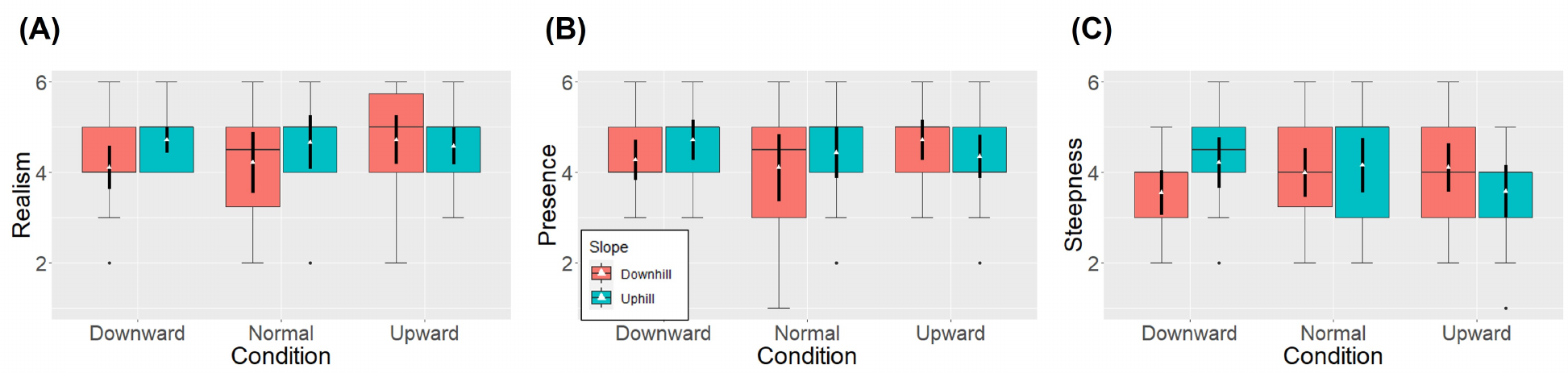}
 \caption{Results of Experiment 3. The white triangle indicates the mean, and the bold line shows the 95\% confidence interval. The error bar indicates the theoretical minimum or maximum value. Regarding a box plot, the middle horizontal lines show the medians, the colored boxes show the upper and lower quartiles, and the dots outside of the error bar show outliers. All vertical axes indicate the rating score for the questionnaire items of (A) realism, (B) presence, and (C) steepness.}
 \label{ALL_XP3}
\end{figure*}

Fig.\ref{ALL_XP3} displays the results of questions about the realism of the walking experience, the steepness of the slope, and the sense of presence in VR. The Wilcoxon signed rank test with Holm correction was conducted for each question score between WIP conditions within each slope condition, and no statistical differences were found. In addition, we conducted the same test between slope conditions within each WIP condition for all questions. As a result, a statistical difference was found in the downward condition for realism ($p=0.049$).

\subsection{Discussion}
\subsubsection{Can pseudo-haptic approach with WIP simulate a virtual slope?}
\textbf{H3-1} was supported by the fact that the speed gain that participants thought was appropriate for the virtual slope was 0.71 $\pm$ 0.07 for uphill and 1.43 $\pm$ 0.25 for downhill in normal (without elastic) condition. This result suggests that the visual representation of relatively slow progress on the uphill and fast progress on the downhill modified the walking load perception in the manner of pseudo-haptics and improved the appropriateness of the slope representation. Moreover, all questionnaire evaluations were positive, suggesting that pseudo-haptics with WIP could improve UX. However, we did not include the condition where the speed gain was 1.0 in the evaluation task (the second part of the experiment) due to the task volume. Future work investigating the effect of different speed gains on the realism or steepness of the slope will more clearly determine the benefits of pseudo-haptics with WIP. Moreover, other future work can investigate the possibility of the pseudo-haptics technique with other VEs, such as mud or icy road.

\subsubsection{Does elastic input with pseudo-haptic approach increase the realism of a virtual slope?}
\textbf{H3-2} was not supported in this experiment: statistical differences were not found between the normal condition and other elastic input conditions. As mentioned above, most of the scores in the questionnaires were positive, and therefore, it is possible that the effect of pseudo-haptics was greater than that of the elastic input on the realism of the slope. Thus, the effect of the elastic input became relatively small, and no statistical difference was observed. Another possibility is that the number of participants was not large enough to find a statistical difference of small effect size. A statistical difference was observed only between uphill and downhill at the downward elastic condition in the results of the "realism" question. However, Figure \ref{ALL_XP3} indicates some tendencies showing that the downward elastic input suited the uphill representation and the upward elastic input was usable for the downhill representation. This would need to be confirmed by future work with more participants.

\section{Conclusion}
In this paper, we revisited the walking-in-place interaction scheme in VEs. We investigated novel methods for its three main components, namely control, input, and output. 
Regarding the control factor, we proposed a novel WIP technique called SHeF-WIP, which utilizes the step height and frequency for virtual speed determination. Furthermore, regarding the input factor, we investigated a vertical elastic input on the feet during WIP. Finally, regarding the output factor, we introduced a pseudo-haptic approach to WIP to simulate walking over virtual slopes.

Interestingly, these contributions affected each other. To discuss the relationship between step height control and elastic input, a higher step height causes a faster virtual speed and a stronger-downward or weaker-upward elastic force. The synergy between elastic input and pseudo-haptic feedback has already been confirmed in some previous studies \cite{Paljic:2004, paljic2004evaluation, Achibet:2014, Achibet:2015}. Moreover, to discuss the relationship between pseudo-haptic feedback and step height and frequency control, SHeF-WIP could reflect the behavioral changes induced by pseudo-haptic feedback on the output speed better than step frequency control only, and this could further improve the realism of VR. 

We conducted three experiments to evaluate the effects of novel interaction schemes on WIP performance and UX.
The results of Experiment 1 suggest that regarding SHeF-WIP control, the participants appropriately changed their WIP strategy using the step height and frequency depending on the virtual walking speed. Furthermore, SHeF-WIP enabled the participants to reach speed increases of up to +80\% compared with GUD-WIP while achieving greater stability and ease. The results of Experiment 2 suggest that an upward-strong elastic input can effectively improve the stability of the SHeF-WIP control compared to the SHeF-WIP without elastic input. However, a strong elastic force may decrease the realism of walking and the usability of the system. The results of Experiment 3 indicate that pseudo-haptic feedback works efficiently in the presence of WIP. The appropriate speed gain was below 1.0 for uphill and above 1.0 for downhill. 

Taken together, our generally positive results suggest that, for future VR applications, there is value in further research into the use of alternative interaction schemes for walking-in-place.

One of the limitations of this paper is that the number of participants might not be enough to observe a statistical difference of a small effect size. As seen in some figures of the results, some results show tendencies of difference. Future studies with a larger number of participants may find the statistical differences for that. Another limitation of this paper is that the gender population was not well-balanced. Several studies show that women tend to rely more on visual information than men (e.g. \cite{tremblay2004gender, nguyen2018individual}). Therefore, especially the results of XP3 may be different between men and women. This should be investigated in future work.
Moreover, it should be also noted that the current elastic feedback implementation does not allow users to physically turn. Future work can improve the elastic feedback system with less restriction on the user's movement.
Lastly, other future work should explore the system in more ecologically valid scenarios.

\section*{Acknowledgments}
This work was partially supported by the MEXT Grant-in-Aid for Scientific Research(S) (19H05661) and Grant-in-Aid for JSPS Fellows (21J12284).

\ifCLASSOPTIONcaptionsoff
  \newpage
\fi



\bibliographystyle{IEEEtran}
\bibliography{main}
%


%

\vspace*{-3\baselineskip}

\begin{IEEEbiography}[{\includegraphics[width=1in,height=1.25in,clip,keepaspectratio]{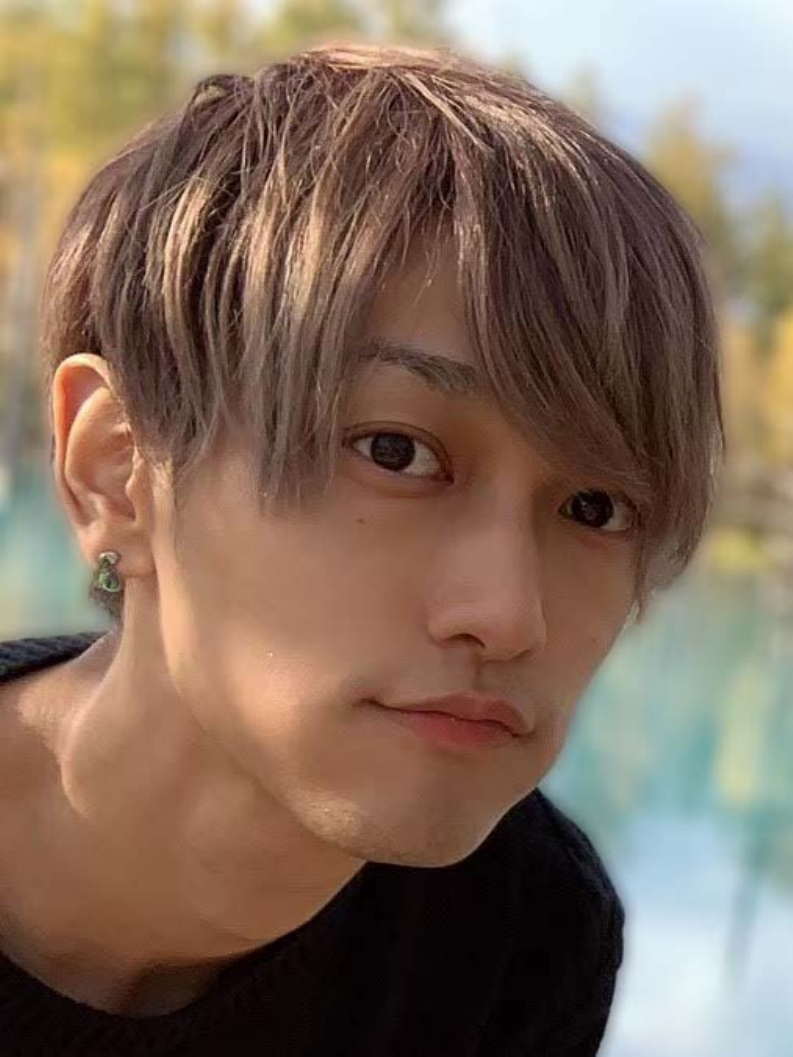}}]{Yutaro Hirao}
received his B.S. and M.S. in engineering from Waseda University (2018 and 2020) in Japan. He is currently working toward
the PhD degree in engineering from the University of Tokyo. His research topics are mainly virtual reality (VR) and cross-modal interaction.
\end{IEEEbiography}

\vspace*{-3.4\baselineskip}

\begin{IEEEbiography}[{\includegraphics[width=1in,height=1.25in,clip,keepaspectratio]{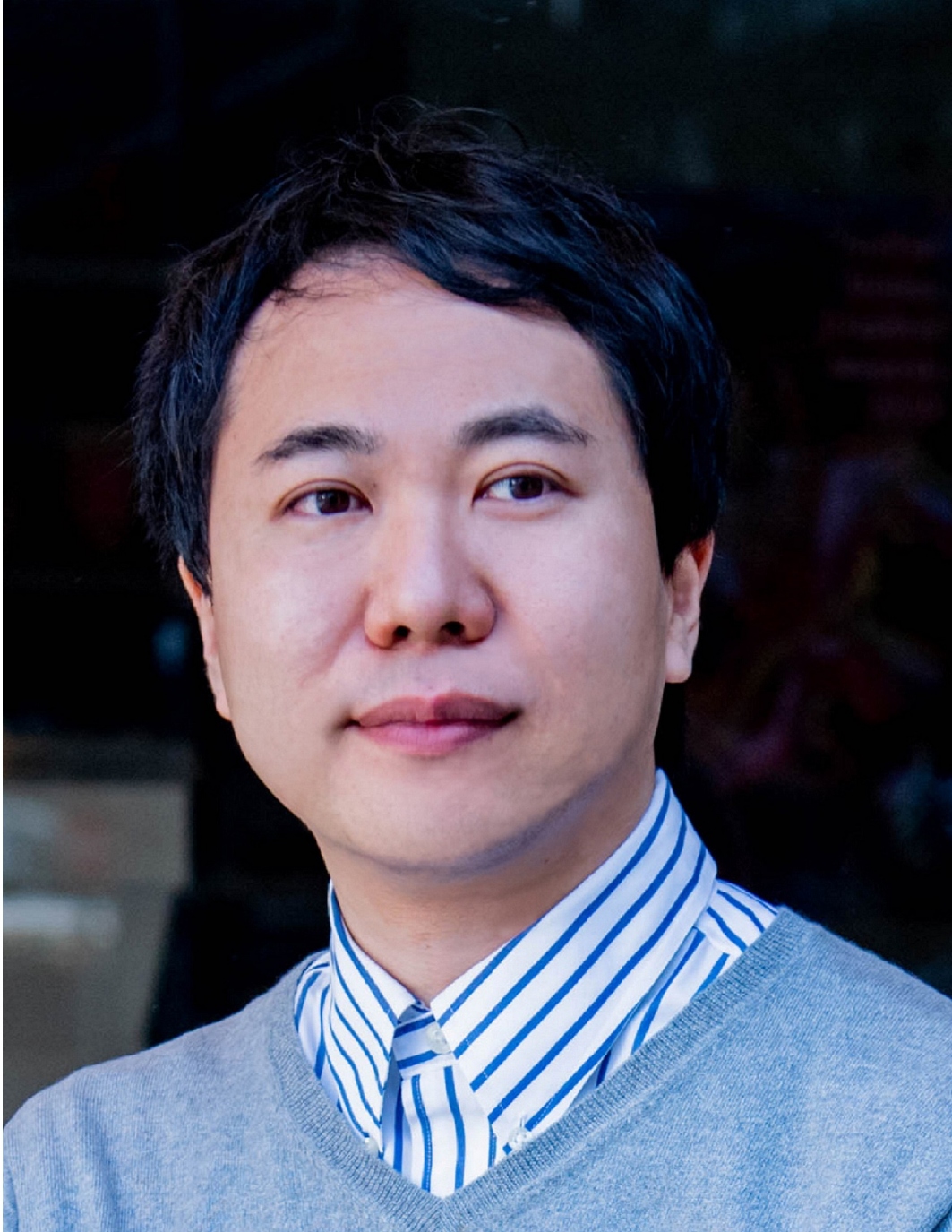}}]{Takuji Narumi}
is an associate professor at the Graduate School of Information Science and Technology, the University of Tokyo. His research interests broadly include perceptual modification and human augmentation with virtual reality and augmented reality technologies. He received BE and ME degree from the University of Tokyo in 2006 and 2008 respectively. He also received his Ph.D. in Engineering from the University of Tokyo in 2011.
\end{IEEEbiography}

\vspace*{-4\baselineskip}

\begin{IEEEbiography}[{\includegraphics[width=1in,height=1.25in,clip,keepaspectratio]{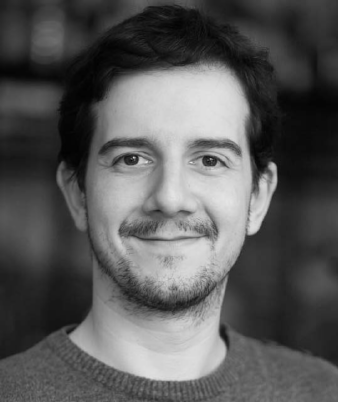}}]{Ferran Argelaguet}
is an Inria research scientist at the Hybrid team (Rennes, France) since 2016. He received his PhD degree from the Universitat Polit\`ecnica de Catalunya (UPC), in Barcelona, Spain in 2011. His main research interests include 3D user interfaces, virtual reality and human-computer interaction. He was program co-chair of the IEEE Virtual Reality and 3D User Interfaces conference track in 2019 and 2020, and the journal track in 2022.
\end{IEEEbiography}

\vspace*{-4\baselineskip}

\begin{IEEEbiography}[{\includegraphics[width=1in,height=1.25in,clip,keepaspectratio]{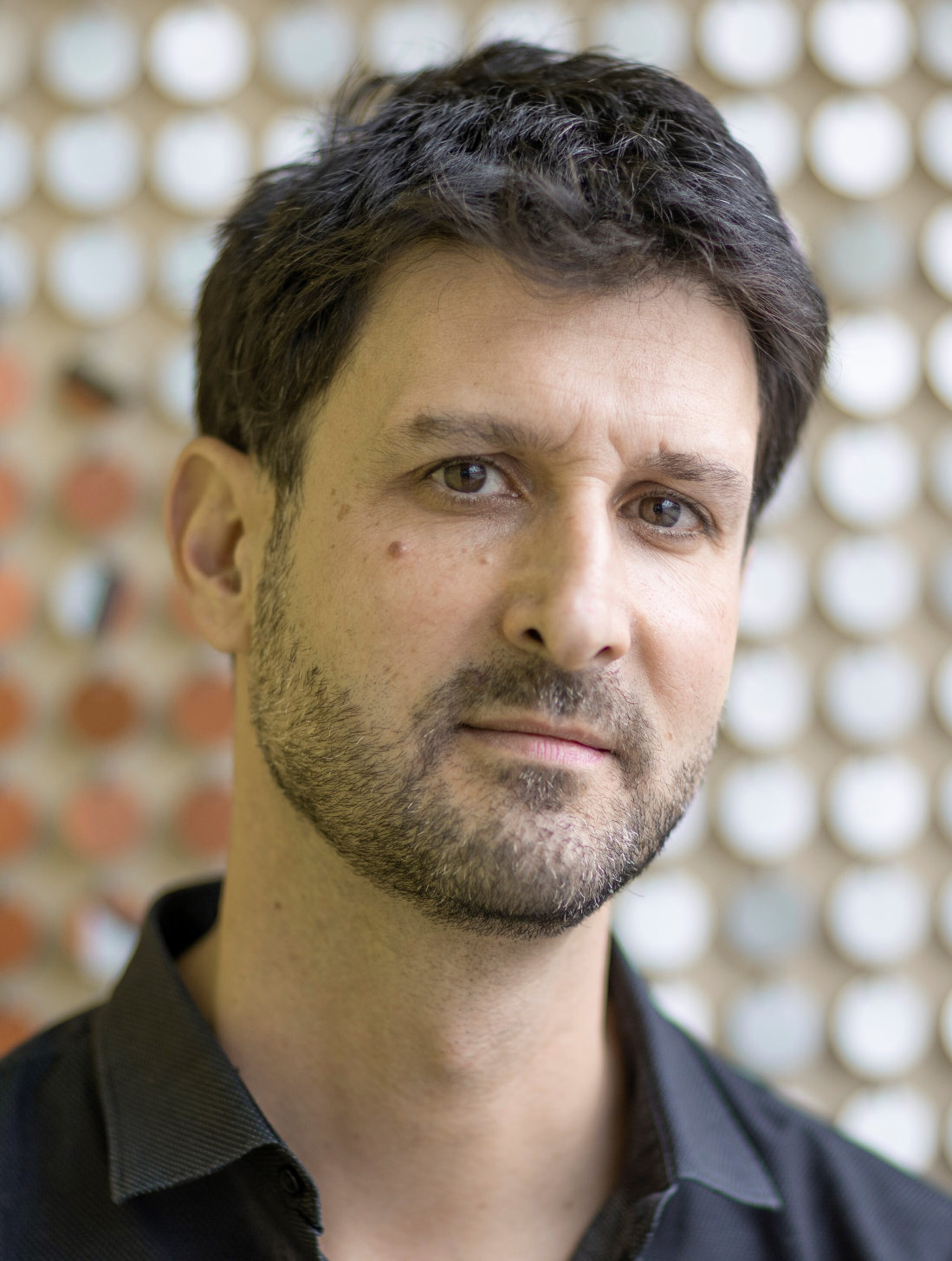}}]{Anatole Lécuyer}
is director of research and head of Hybrid team at Inria, Renne, France. He is currently Associate Editor of IEEE Transactions on Visualization and Computer Graphics, Frontiers in Virtual Reality and Presence. He was Program Chair of IEEE VR 2015-2016 and General Chair of IEEE ISMAR 2017. Anatole Lécuyer obtained the IEEE VGTC Technical Achievement Award in Virtual/Augmented Reality in 2019.
\end{IEEEbiography}

\vspace*{-3\baselineskip}







\end{document}